# The Small Unit Cell Reconstructions of SrTiO$_3$ (111)

Laurence D. Marks[1], Ann N. Chiaramonti[1], Fabien Tran[2] and Peter Blaha[2]

[1]Institute for Catalysis in Energy Processes and Department of Materials Science and Engineering, Northwestern University, Evanston, IL 60208, USA
[2]Institute of Materials Chemistry, Vienna University of Technology, Getreidemarkt 9/165-TC, A-1060 Vienna, Austria

**Abstract**

We analyze the basic structural units of simple reconstructions of the (111) surface of SrTiO$_3$ using density functional calculations. The prime focus is to answer three questions: what is the most appropriate functional to use; how accurate are the energies; what are the dominant low-energy structures and where do they lie on the surface phase diagram. Using test calculations of representative small molecules we compare conventional GGA with higher-order methods such as the TPSS meta-GGA and on-site hybrid methods PBE0 and TPSSh, the later being the most accurate. There are large effects due to reduction of the metal d oxygen sp hybridization when using the hybrid methods which are equivalent to a dynamical GGA+U, which leads to rather substantial improvements in the atomization energies of simple calibration molecules, even though the d-electron density for titanium compounds is rather small. By comparing the errors of the different methods we are able to generate an estimate of the theoretical error, which is about 0.25eV per 1x1 unit cell, with changes of 0.5-1.0 eV per 1x1 cell with the more accurate method relative to conventional GGA. An analysis of the plausible structures reveals an unusual low-energy TiO$_2$-rich configuration with an unexpected distorted trigonal biprismatic structure. This structure can act as a template for layers of either TiO or Ti$_2$O$_3$, consistent with experimental results as well as, in principle, Magnelli phases. The results also suggest that both the fracture surface and the stoichiometric SrTiO$_3$ (111) surface should spontaneously disproportionate into SrO and TiO$_2$ rich domains, and show that there are still surprises to be found for polar oxide surfaces.

Keywords: DFT, Polar Oxide Surfaces, Surface Structure

**Introduction**

Oxides are ubiquitous as they are present on earth and in every soil and sediment as well as in aerosols, aquatic biota, and waste streams. They come from a variety of sources, both natural and anthropogenic. For many bulk oxides the crystal structures are well established, and for most systems the thermodynamics are well documented from experiments and theoretical calculations. At the surface much less is understood. While one can easily perform theoretical calculations on simple bulk-like (e.g. 1x1) terminations, the actual thermodynamically stable surface structures are often larger and more complicated. Until these surface structures have been unambiguously experimentally determined, the problem can be confused. Even for such a simple system as the (100) surface of the archetypal perovskite strontium titanate, there are at least six different experimentally determined surface reconstructions in addition to the 1x1 bulk terminations. Not all of these structures have yet been solved at the atomic level, and to date there is not convincing agreement between experimental observations and theoretical analyses as to which surface structures should be stable under what conditions.

More complicated yet is the problem of the so-called polar surfaces. These are surfaces where a simple Gibbs truncation to form a 1x1 structure leads, in a fully ionic model, to a surface with a nett charge and/or an unbalanced dipole which would lead to an infinite surface energy. For an isolated surface in vacuum or a gas, except for special cases where charge has deliberately been created (for instance with a Van Der Graaff generator), the nett charge of a true surface is always zero and the energy will never be infinite. There has been extensive discussion of the mechanisms of "charge compensation" for oxide surfaces in the literature (see for instance [1,2] and references therein), but typically they assume a fully ionic model, which is not correct in the case of many transition metal oxides. A nice, very recent demonstration of this is the analysis by Raebiger et al of charge self-regulation in transition metal ions in semiconductors [3]. An alternative concept is to consider what we will refer to as "valence compensation". Attributing a nominal charge equal to the valence of each atom type, if this is nett zero, is likely to give an insulator with a relatively large band-gap; if not there will probably be either n-type or p-type states. Since the band gap in many oxides is large in most cases the extra energy associated with creating a hole in the valence band would make the surface a stronger oxidant than molecular oxygen; an electron in the conduction band well capable of reducing water. This is physically unlikely so in many cases the polar surfaces will rearrange to form more stable and redox neutral configurations.

A very specific case is the $SrTiO_3$ (111) surface. A wide range of reconstructions have been observed depending upon the annealing time (on the scale of hours), the temperature, the oxygen partial pressure, and whether the specimens were ion beam sputtered before analysis including (1x1) [4-6], (9/5 x 9/5) [7-9], ($\sqrt{7}$x$\sqrt{7}$ R19.1°) [10], (3x3) [7-9], ($\sqrt{13}$x$\sqrt{13}$ R13.9°) [10], (4x4) [7-9], (5x5) [9], (6x6) [7-9] as well as a TiO overgrowth under highly reducing conditions [9]. While the approximate chemistry and morphology of the surfaces is relatively well characterized, nobody has yet proposed and verified an atomic-level structure solution which includes locating the three-dimensional positions of all the atoms in the reconstructed surface selvedge and verifying the result with structure refinement and/or simulation against scattering data. There is only one theoretical study in the literature using a semi empirical Hartree-Fock method [11,12], and this considered a few relatively simple 1x1 and 2x1 structures, not the much larger reconstructed cells observed experimentally as well as some other possibilities that we will discuss later.

It has now become almost conventional when proposing a model for a surface reconstruction to perform a density functional theory (DFT) calculation. What one wants to know is whether the proposed positions are plausible, i.e. the difference between them and refined DFT positions is not too large, as well as whether energetically the structure is plausible. For this one needs as background information to have answered three fundamental questions:

1) What is the most appropriate DFT functional to use; there are many in the literature.
2) What are the errors in the energies? These numbers are rarely analyzed or published and from an experimental viewpoint a measurement without errors is marginal. Obviously only with knowledge of the errors in the energies can one determine if a structure is plausible.
3) What are the basic simple structures against which one wants to compare a reconstruction? Since often reconstructions are variants/superstructures based upon simple 1x1 units, this information is also needed *a-priori* to aid in solving reconstructions, particularly ones with large unit cells such as observed for the (111) surface of $SrTiO_3$.

The focus of this paper is to provide some answers to the three questions above for the (111) surface of $SrTiO_3$; the more complicated nxn reconstructions is a topic of further publications. The structure of the paper is as follows. After a brief, technical description of the parameters used, we turn to an analysis of what functional to use. Our approach is to analyze the atomization energies of some representative small molecules, as a reasonable model for changes in bonding at a surface (which dominate the relative surface energies). We point out that even though titanium has a rather small d-electron population it is important to consider higher-order methods which compensate for inaccuracies in the d-electron exchange-correlation potential. An efficient method of doing this is an on-site hybrid-DFT (PBE0) method, where one adds a fraction of exact exchange (inside sphere and only for selected electrons) and which is similar to a GGA+U method but with a dynamically calculated Hubbard U term. The best results are obtained using a meta-GGA combined with a hybrid functional. By comparing the errors for different functionals for these known cases, we can estimate that the theoretical error for the surface energies in terms of the difference in the results for different functionals. We then turn to the basic structures, including a number of not-so-simple 1x1 reconstructions on the TiOx-rich part of the surface phase diagram. We find a surprisingly low-energy structure with an unexpected structural motif of distorted trigonal bi-prisms, rather than octahedral, tetrahedral or 5-fold units. This structure acts as a ready template for additional growth of TiO or $Ti_2O_3$ layers, consistent with experimental results. These results are combined into a partial phase diagram of the simple structures.

**Methods**

For the DFT calculations the all-electron Wien2k code [13] with an augmented plane wave (APW) basis set was employed. For reference, technical parameters were: atomic sphere sizes (RMT's) of 2.36, 1.72 and 1.54 a.u. for Sr, Ti and O respectively, a Fourier series cutoff of GMAX = 21.6 for the charge density and potential, and a wavefunction cutoff (defined as product of the smallest RMT times the largest K in the plane wave expansion) of RKMAX = 6.12. The Brillouin zone sampling was 5x5x1 for the 1x1 cell, scaled for the larger cells to retain approximately the same density of points in reciprocal space. A small (0.0018 Rydberg) temperature factor corresponding to the Fermi-Dirac occupation at room temperature was used; this had little effect since most of the relevant structures were insulators. The separation between the two surfaces was at least 1nm, with total slab sizes of 2-2.5nm. Tests indicated that the intrinsic numerical errors such as convergence as a function of reciprocal-space sampling were < 0.01eV per 1x1 unit cell, which is much smaller than the variations with different functionals as detailed below. In all cases the surface energies were determined by subtracting the appropriate energies for bulk $SrTiO_3$, SrO or $TiO_2$, calculated in larger supercells (for instance a hexagonal cell with a=[110] and c=[111] for $SrTiO_3$) with matching technical parameters to minimize numerical errors. Although initial calculations were performed using the conventional generalized gradient approximation (GGA) as defined by the PBE functional [14], for reasons that we will discuss in more detail below for the final calculations we used an on-site Hartree-Fock hybrid method [15], namely the PBE0 functional [16,17] for the exchange-correlation potential, while the exchange-correlation energy is taken from the meta-GGA TPSS functional [18], equivalent to an on-site TPSSh method [19]. On-site means in this context that the exact-exchange (Hartree-Fock) part is calculated only for selected electrons (Ti-3d) inside the corresponding Ti sphere [15], thus keeping the numerical effort quite small. For reference, both PBE and the TPSS calculations were performed with the PBE minimized lattice parameters, while the PBE0 and TPSSh results were obtained with the refined lattice parameters for PBE0.

**Choice of DFT Method**

While DFT calculations to accompany experimental surface structure analyses have become common, one has to be careful when performing such correlations. Particularly in the case of oxides, the method must do a reasonable job of taking account of not simply the bulk bonding, but how the covalent and ionic bonding changes at a surface as well as the long-range surface energy contributions. The most common functional used is the PBE GGA [14], which while it often gives very good results, still has some problems particularly for energies. For instance, it is now well established that it severely underestimates surface energies [20-22]. In addition, for many bulk transition element oxides there is too much hybridization between the metal 3d-electrons and the oxygen 2sp, and the classic method for correcting this is what has become known as the LDA+U method [23,24]. In general this increases the ionicity of the bonding, which can be badly underestimated otherwise. The method requires a number for the value of the Hubbard U that is difficult to determine independently and will depend upon the local environment so is not a global parameter. While one would not normally consider $SrTiO_3$ or insulating compounds containing Ti as cases where one has to use this method (because the d-electron density is small), even here there is a noticeable hybridization between the oxygen 2sp levels and the Ti d-levels.

A recently developed alternative to the LDA+U method, which involves less in the way of arbitrary parameters, is to use an on-site hybrid [15] based upon an approach such as the PBE0 functional [16,17]. One adds a small component of exact-exchange for the relevant Kohn-Sham orbitals (d-electrons only here), which can be calculated rather simply within the muffin tin radii of an APW method. It has been shown [15] that this is similar to a LDA/GGA+U method, but with a U value that is dynamically calculated and will vary with environment. Strictly speaking the exact amount of exact-exchange is system dependent but in many cases a fixed value works, for instance 0.25 for the PBE0 functional (see also the discussion). This is important because it removes the issue of what value of U is relevant for a surface atom as against a bulk atom since they are different; one uses a fixed value for the amount of exact exchange and the calculation automatically adjusts the effective U value depending upon the environment. In addition, a method for calculating the forces has recently been developed [25] so the on-site method can be applied for a full structural minimization with at most a 10% overhead relative to a conventional GGA calculation.

While the on-site hybrid approach appears to be better for energies (see below), we can go a bit further. As mentioned above, the PBE functional gives poor surface energies. A much better method for this is the TPSS meta-GGA functional [18] which includes beside the gradient of the density also the kinetic energy-density in the functional form which is known to match quite well the long-range jellium surface energies which are believed to be an issue with PBE [20-22]. TPSS also gives much better atomization energies for molecules. It corrects, for instance, the overestimation of the atomization energy for $O_2$ which is ~6.5eV with PBE whereas TPSS gives ~5.3eV, a value that is closer to the experimental result of 5.12 eV (e.g. [26]). While it is hard to implement this functional in a fully self-consistent fashion it is common practice to use the electron density corresponding to e.g. a PBE potential and includes only the exchange-correlation energy contribution and this is known to give quite good results (as suggested first in the original publication [18]) A combination of full TPSS and PBE0 is called the TPSSh functional; the on-site version will be referred to as an on-site TPSSh method [19]. (The original papers used the TPSSh functional with a small amount of exact exchange, 0.1; for the on-site implementation the same value as for PBE0 of 0.25 works better at least for the systems studied herein.)

For completeness, we also tested a recent one-parameter optimization of TPSS [27]. While this gave slightly better atomization energies, the difference was small and slightly worse for the disproportionate reaction so it will not be used here.

To verify that this method is an improvement, the atomization energies of a number of small Ti+O molecules for which experimental data are available (see [28,29] and references therein) were calculated, with the experimental values corrected for the zero-point energies. For completeness and to understand the limits of applicability, the data for TiN was also included, (see [30] and references therein) where the calculated values particularly with TPSSh are not as accurate as they are for the other compounds; the method is not perfect. The indirect DFT bandgap of bulk $SrTiO_3$, which is of course not the same as the true bandgap of 3.77eV [31] .but representative of it, was also examined. Since for TPSS and TPSSh functionals we do not have a self-consistent electronic structure, the gap was estimated from the total energies by modifying the Fermi-Dirac occupancy so that N-$\delta$ electrons were below the valence-band edge and $\delta$ above the conduction-band edge, and iterating to self consistency. (For completeness, this gave identical values for the band-gap for PBE as one would obtain from the difference between the lowest unoccupied highest occupied eigenvalues). In addition we looked at the disproportionation reaction for which the experimental value is 1.425eV [32]:

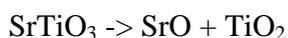

$SrTiO_3 \rightarrow SrO + TiO_2$

Table 1 summarizes the results, with the common hybrid method B3PW91 [33] (on-site form) included for reference. We can rank the accuracy of the methods as PBE << TPSS ~ PBE0 << TPSSh, except for TiN where PBE0 is best, although even with TPSSh the absolute errors in the energies are still significant. We also show in the table the results for an effective Hubbard-U parameter $U_{eff}$ obtained from an L2 fit of the orbital potential versus a conventional U method with J=0, which in this case works well with an RMS error of ~0.15 although we should caution that in some other cases (e.g. an isolated Ti atom, RMS error ~0.7) it is not a good description. As one would expect the $U_{eff}$ value increases as the electron density around the Ti atom decreases, i.e. the shielding of the d-electrons decreases.

Perhaps surprising is that the simple PBE0 method, which mixes in 25% of exact Hartree-Fock exchange, has such a substantial effect since the total d-electron density is only about one electron per titanium atom. The correction to the exchange energy and potential shifts up the energy of the d-electrons, reducing the degree of hybridization of the oxygen 2sp levels with the titanium 3d and as a consequence making the bonding more ionic. One can estimate the ionic charges using Bader's "atoms in molecules (AIM)" theory [34], , which for bulk $TiO_2$ give a nominal Ti charge of +2.28 with PBE versus +2.43 with PBE0. For completeness we note that the fact that conventional PBE calculations underestimate the ionicity of systems containing Ti is supported by other experimental evidence such as charge density data [35].

The calculations of the energies for the small molecules is not perfect, neither will be the energies for surfaces. We need to have an estimate of the error to be able to determine anything. From Table 1 a reasonable estimate is to take $\sigma = |E_{TPSSh} - E_{PBE}|/3$.

**Calculated T=0K Energetics**

Figure 1 summarizes the compositions of the different structures in terms of a surface analogue of a conventional three-component phase plot. While the energies for the (less accurate) PBE method are qualitatively similar to what is found with the TPSSh method, because the later appear to be much more accurate only those are reported. In detail, the specific structures were:

1) Models 1 & 2, a simple bulk-terminated (111) SrTiO$_3$ 1x1 surface, both Ti (Model 1, Figure 2a) and SrO$_3$ terminations (Model 2, Figure 2b). Neither of these are stoichiometric, the Ti termination being Ti rich and oxygen deficient (n-type) whereas the SrO$_3$ termination is oxygen and strontium rich (p-type). Contrary to the earlier semi-empirical calculations we find that both of these are metallic in nature as one would expect.

2) Model 3, a valence compensated SrO termination (SrO rich) which corresponds to a c2x1 reconstruction with oxygen vacancies in the surface as shown in Figure 2c, similar to that analyzed by Pojani et al [11,12].

3) Model 4, a valence compensated 1x1 TiO termination (TiO$_2$ rich) with a tetrahedrally coordinated surface Ti atom with an oxygen atom above it, shown in Figure 2d, similar to that analyzed by Pojani et al [11,12].

4) Models 5 and 6, two possible valence compensated SrTiO$_3$ termination with half the Ti atoms missing in the outermost Ti layer, with 5 shown in Figure 2e (a 2x1 cell) and 6, in Figure 2f (a 2x2 cell).

5) Model 7, a TiO$_2$ rich valence compensated 1x1 reconstruction with a Sr vacancy in the 2$^{nd}$ layer, Figure 2g.

6) Model 8, (Figure 2h) a TiO$_2$ rich valence compensated 1x1 reconstruction where in addition to a Ti atom in the normal position at the surface, another is placed above one of the Ti atoms in the layer below and the structure is terminated with a layer of oxygen atoms. This structure has a somewhat unusual surface co-ordination for the Ti, distorted trigonal biprismatic.

7) Models 9 and 10, two more TiO$_x$ rich structures based upon adding layers to Model 8. In Model 9 (Figure 2i, left) the additional layer has a stoichiometry of Ti$_2$O$_3$ which for Model 10 (Figure 2i right) it is 3(TiO), and both are 1x1 cells. For the Ti$_2$O$_3$ structure two of the three possible 3-fold sites above the terminal oxygen layer of Model 8 are occupied, for the TiO structure all three.

Refined positions for these different models are available as conventional crystallographic cif files [36]. An analysis of aspects of the lower-energy structures will be given later in the discussion.

Two additional combinations of structures are important. The first is what we will call the "Gibbs Surface", generated by performing a planar cut of the bulk. This will be the average of the simple 1x1 terminations of Model 1 and 2. The second is the "Fracture Surface" which we define as the lowest energy combination of two structures which with minimal relaxations can be combined to form a perfect bulk. For this system this is the average of the oxygen vacancy Model 3 and the oxygen adatom Model 4. For completeness, note that the energy of the Fracture Surface will be lower than the true fracture energy of the sample on a (111) plane.

Shown in Table 2 are the surface energies per 1x1 unit cell for all the structures analyzed and the functionals PBE, PBE0, TPSS and TPSSh, referenced to bulk $SrTiO_3$ plus the relevant fractions of bulk $TiO_2$ and $O_2$. There are several ways to interpret the data. A conventional method would be to plot the energies versus a relevant chemical potential, for instance that of $TiO_2$ taking the chemical potential of $SrTiO_2$ as a reference (via the disproportionation energy) to determine the chemical potential of SrO. Unfortunately it is hard to justify this in terms of the experiments. The system is not in equilibrium with any source of $TiO_2$ to fix the chemical potential. While in principle it is in equilibrium with bulk vacancy clusters of SrO and $TiO_2$, at the temperatures of interest experimentally the kinetics are sluggish. Indeed, the surface structures depend upon the history of treatment of the sample, particular annealing time and oxygen pressure [7,8], We note that even with respect to oxygen we should not assume global equilibrium, rather only local equilibrium similar to the $TiO_2$ (001) surface [37].

What is more relevant is to consider the energetics for a fixed surface excess of (for instance) $TiO_2$, and consider what combination of structures are thermodynamically of lowest energy. To do this we set the chemical potential of both $SrTiO_3$ and $TiO_2$ as zero and plot the energies versus composition as shown in Figure 3. One then does a conventional convex-hull construction, connecting all points on an energy-composition diagram and takes the lowest energy combination; for structures containing either excess or less oxygen one would need to correct for the oxygen chemical potential for specific temperatures and pressures. The combination of phases (one or two) can then be generated by a conventional lever-law at any composition. It should be noted that as constructed, if we for instance change the chemical potential of $TiO_2$ this simply rotates the diagram and has no effect upon the relative phase composition for a fixed composition.

The results are shown in Figure 3 for the valence compensated compositions relative to the SrO-rich Model 3 for both PBE and TPSSh. With an estimate of the error as mentioned earlier, excluding the non-stoichiometric Models 1 & 2 (to avoid issues with the oxygen molecule bonding), we can estimate a value of 0.25eV/1x1 unit cell which has been used in the Figure. While it is clear that if one compares relative energies the two different functionals give qualitatively comparable results, there are clearly rather large differences in the quantitative numbers (even when, as in this Figure, an offset of 0.84eV has been eliminated).

The analysis indicates that (of the structures considered) the stable phases are the SrO-rich Model 3, the structure with prismatic Ti of Model 8 close to zero oxygen chemical potential; under highly oxidizing conditions the two bulk truncated 1x1 terminations and under highly reducing conditions the $Ti_2O_3$ or TiO overgrowths. This is summarized in Figure 4, excluding to two highly $TiO_x$ rich structures. For completeness, we can only base an analysis on the phases considered, and others might be relevant. Note that the results indicate that the stoichiometric SrTiO$_3$ reconstructions as well as the fracture surface should disproportionate, with a 1.5σ (90%) confidence.

Some analysis of the effect of moving from the PBE functional to PBE0 and TPSSh is appropriate at this point. Several observations can be made:

1) The surface energies increase in the order PBE < PBE0~TPSS < TPSSh. The increase with PBE0 is in part because the system overall has become more ionic, and because the PBE method will be overestimating the possible covalent stabilization of the surface. The increase in the energy with the use of the meta-GGA TPSS is expected, and is one of the well-established problems with PBE. As a first approximation the ratio of the surface energies $E_{TPSSh}/E_{PBE}$ is ~1.3 with an accuracy of ~0.1eV per 1x1 unit cell.

2) The change in the enthalpies is not independent of which surface structure one has, and varies by as much as 0.5eV between them. In general the increase is larger the higher the density of exposed titanium atoms at the surface, or the electron density in the d-levels as is the case for the n-type Ti terminated structure. This is exactly what one would expect for the increased ionicity.

3) While many features of Figure 3 do not depend upon which functional is used, some do and in a predictable fashion. The largest effect is seen when one compares model 7, a 1x1 with a missing Sr atom in the second layer, to the biprismatic Model 8. This change is because model 7 has an exposed Ti atom which is much more ionic when a better account is taken of the Ti-d, O-sp hybridization.

4) As one would expect, the effective U value ($U_0$=4.93 eV in the bulk) increases at the surface by 0.2-0.5 eV with the exception of Model 2 where there is no change.

**Discussion**

We will discuss first the three questions which were posed in the introduction. There is a clear improvement when going from the standard GGA based upon the PBE method to the meta-GGA as well as the hybrid form, even the comparatively simple on-site implementation that we have used here. This is consistent with several recent analyses in the literature, e.g. [30,38-41]. For the systems studied herein one of the largest issues is the hybridization between the cation d-levels and the anion sp-levels, and this is improved by using a better form for the exchange potential of the d-levels. One issue is exactly how much of exact-exchange should be used. As analyzed in the original paper by Perdew, Ernzerhof and Burke [16] a value of 0.25 is close to optimum, but may not be best in all possible cases. For instance, for the NiO (111) surface which we will discuss elsewhere, a value of 0.20 appears to be rather better for the Ni atoms, matching both the bulk properties as well as simple test molecules such as NiO, Ni(CO)$_4$ and Ni(CO)$_3$. What does appear to be the case is that this method has large advantages over conventional LDA+U as an auto-adjusting U method. While not easy, it should be possible to measure the relative surface energies experimentally particularly as a function of oxygen chemical potential to rather better than this level, so there may be some direct tests available.

The method we have used to estimate the accuracy, i.e. use the errors for small calibration molecules with different functionals to estimate an error based upon the difference between two (or more in principle) is not perfect, but reasonable; it has been used in a slightly different form earlier with LDA and PW91 functionals [42]. In principle one might be able to do better with a Bayesian method as suggested by Mortensen et al [43]. The accuracy that we argue the calculations have, ~0.25 eV per 1x1 unit cell, is probably the best that can be achieved with current DFT methods. As such these are still some distance from the levels where phonon entropy effects will be important at low temperatures, although they might start to be important at the higher temperatures used in experiments – the kinetics of equilibration for many oxides are slow except at quite elevated temperatures.

The majority of the structures are relatively simple, and most of the energies are not too far from what one might expect based upon basic inorganic chemistry. In general the larger the number of "bonds" between cation and anions, the lower the energy. The very low energy triangular prismatic Ti co-ordination is a bit of a surprise since this is not a standard co-ordination for Ti compounds. In hindsight it can be rationalized as due to a large co-ordination for both the cations and anions; this over-rules the standard co-ordination chemistry drives to form more octahedral structures. While it would be nice to believe that we understand enough about polar oxide surfaces that there are no more surprises to be come, this strongly suggests that we do not.

The two reduced structures which have either a $Ti_2O_3$ coverage or a TiO coverage follow rather naturally as co-ordinations where the oxygen sublattice is preserved, and the only difference is which sites are filled by the cations, a common occurrence. It is worth commenting that both of these will show a modulation equal to the [110] spacing of $SrTiO_3$, 5.52 Angstroms, if imaged by STM. A slightly larger spacing of 5.9 Angstroms has been reported in a small region [9]. This was interpreted as a octapolar reconstruction of a TiO overlayer and while this is reasonable, it might also have been due to one of these overlayers; more work would be required to determine this. For certain the formation of a TiO phase under highly reducing conditions is consistent with our structures.

The thermodynamics of the $SrTiO_3$ (111) surface are more complicated than what we have considered; experimentally a range of nxn reconstructions are found, exactly which forms depending upon the details of the sample preparation. The available Auger data indicates that these are all rich in both Ti and O relative to the fracture surface. In addition, they all show a dominantly strong diffraction spots at, in terms of the 1x1 lattice, a location of approximately (5/3,0) which *none* of the structures described herein come close to reproducing. The structure of these other reconstructions using both experiments and calculations as well as a more complete thermodynamics is a topic of further publications.

**Acknowledgements**

LDM acknowledges support from the DOE on grant number DE-FG02-01ER45945/A007 and ANC acknowledges support from the DOE on grant number DE-FG02-03ER15457. PB and FT acknowledges support from the Austrian Science Foundation, project P20271-N17.

**Tables**

|        | TiO  | $TiO_2$ | $Ti_2O_3$ | TiN   | Reaction | Gap  |
|--------|------|------|-----------|-------|----------|------|
| PBE    | 0.60 | 0.58 | 0.31      | 0.34  | 0.21     | 1.83 |
| B3PW91 | 0.45 | 0.50 | 0.28      | 0.15  | 0.46     | 2.30 |
| TPSS   | 0.46 | 0.40 | 0.17      | 0.09  | 0.05     | 2.24 |
| PBE0   | 0.32 | 0.43 | 0.20      | 0.12  | 0.22     | 2.15 |
| TPSSh  | 0.13 | 0.19 | 0.03      | -0.21 | 0.08     | 2.40 |
| U      | 4.93 | 5.87 | 4.85,5.61 | 5.53  |          |      |

Table 1: Errors in the atomization energies per atom of some small Ti+O as well as TiN test molecules, the effective values of the L2 fit Hubbard U constant as well as the error in the energy for the disproportionation and the indirect band gap, all in eV. For $Ti_2O_3$ the smaller value of U is for the 2-fold coordinated Ti, the larger for the 3-fold coordinated.(For reference, the experimental band-gap is 3.77eV [31] and the energy for the disproportionation reaction 1.425eV [32]

| Model | PBE   | TPSS  | PBE0  | TPSSh | Comp | Layers | Atoms | dU   | Oxid | Symm |
|-------|-------|-------|-------|-------|------|--------|-------|------|------|------|
| 1     | 7.11  | 7.26  | 7.98  | 8.37  | 0.5  | 13     | 31    | 0.43 | -0.5 | p3m1 |
| 2     | 3.05  | 3.71  | 3.22  | 3.92  | -0.5 | 13     | 34    | 0.00 | 0.5  | p3m1 |
| 3     | 2.37  | 2.63  | 2.64  | 3.00  | -0.5 | 13     | 32    | 0.26 | 0    | cm   |
| 4     | 2.57  | 2.88  | 2.84  | 3.41  | 0.5  | 15     | 38    | 0.43 | 0    | p3m1 |
| 5     | 2.94  | 3.21  | 3.39  | 3.78  | 0    | 11     | 50    | 0.17 | 0    | pm   |
| 6     | 3.35  | 3.68  | 3.89  | 4.37  | 0    | 11     | 100   | 0.43 | 0    | p3m1 |
| 7     | 3.02  | 3.34  | 3.51  | 4.05  | 1.5  | 15     | 34    | 0.51 | 0    | p3m1 |
| 8     | 1.66  | 1.76  | 2.00  | 2.31  | 1.5  | 15     | 44    | 0.17 | 0    | p3   |
| 9     | 6.17  | 5.73  | 7.58  | 7.56  | 3.5  | 17     | 54    | 0.26 | -0.5 | p3   |
| 10    | 14.91 | 13.51 | 17.62 | 16.77 | 4.5  | 17     | 56    | 0.43 | -1.5 | p3   |
| G     | 5.08  | 5.49  | 5.60  | 6.15  | 0    |        |       |      | 0    |      |
| F     | 2.47  | 2.76  | 2.74  | 3.20  | 0    |        |       |      | 0    |      |

Table 2: Surface energies per 1x1 unit cell in eV as well as other information for the different models considered, for models 1 and 2 at zero oxygen chemical potential for an $O_2$ atomization energy of 5.35eV. The composition (Comp) is the number of surface excess units of $TiO_2$ less the number of surface excess SrO per 1x1 unit cell. Also shown in the Table is the number of layers, atoms, the increase in the effective U value ($U_0$=4.93 eV in the bulk) at the surface in eV, the oxidation states relative to $1/2O_2$ and the two-dimensional symmetry.

**Figure Captions**

Figure 1: Surface analogue of a three-component phase plot. The compositions indicated correspond to monolayer excesses of the corresponding species with a stoichiometric SrTiO3 marked by the cross. The various models are marked by numbers. The horizontal line is the locus of valence compensated structures; above are oxidized, below reduced.

Figure 2a: (Color online) Model 1, non valence-neutral Ti terminated surface in top (left) and side (right). The top Ti atoms are dark brown.

Figure 2b. Model 2, non valence-neutral SrO3 terminated surface, top (left) and side (right). The outermost Sr atoms are dark brown.

Figure 2c: Top (left) and side (right) view of Model 3, a c2x1 Valence compensated surface with 2 O vacancies per surface cell

Figure 2d: Top (left) and side (right) view of Model 4, valence compensated TiO terminated surface with the additional oxygen atom (dark blue)

Figure 2e: Top (left) and side (right) view of Model 5, valence compensated 2x1 Ti terminated surface with 2 Ti vacancies per unit cell

Figure 2f: Top (left) and side (right) view of Model 6, valence compensated Ti terminated surface with 50% Ti vacancies

Figure 2g: Top (left) and side (right) view of Model 7, valence compensated 1x1 cell with a Sr vacancy in the second layer.

Figure 2h: Top (left) and side (right) view of Model 8, a valence compensated $TiO_2$ rich 1x1 cell with triangular-prismatic co-ordination.

Figure 2i: Side views of Model 9 (left), a $Ti_2O_3$ layer on top of Model 8 and a 3(TiO) layer for Model 10 (right).

Figure 3: Plot of the energies for the different valence compensated structures relative to the SrO rich Model 3 as labeled, versus stoichiometry with an 0.25eV error bar included, labeled with the relevant model; solid for TPSSh and empty for PBE. Also shown is the fracture surface energy (F), and the Gibbs surface energy (G).

Figure 4: Surface phase diagram, based upon the structures considered only, excluding the more reduced $Ti_2O_3$ and TiO overgrowths.

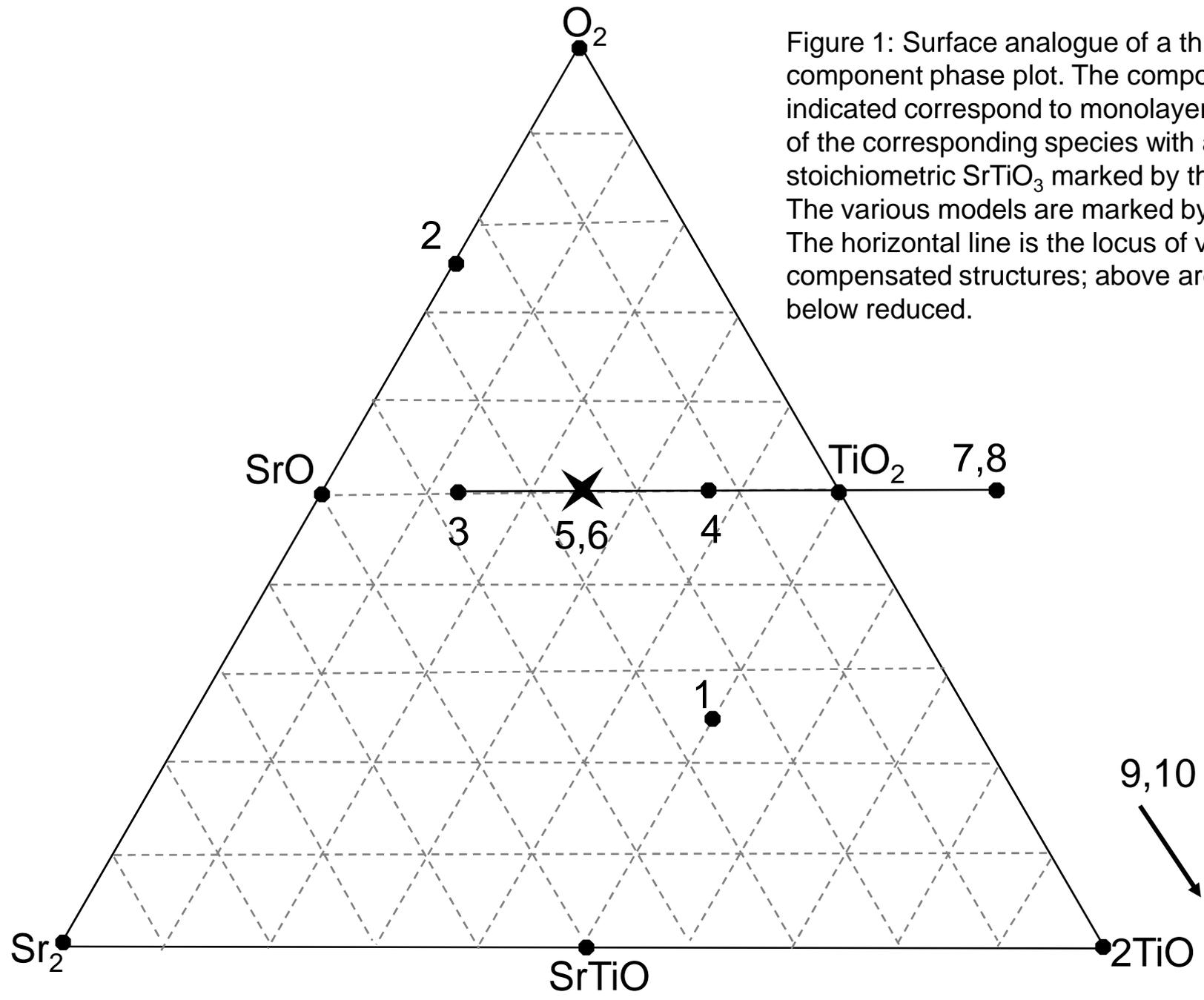

Figure 1: Surface analogue of a three-component phase plot. The compositions indicated correspond to monolayer excesses of the corresponding species with a stoichiometric $SrTiO_3$ marked by the cross. The various models are marked by numbers. The horizontal line is the locus of valence compensated structures; above are oxidized, below reduced.

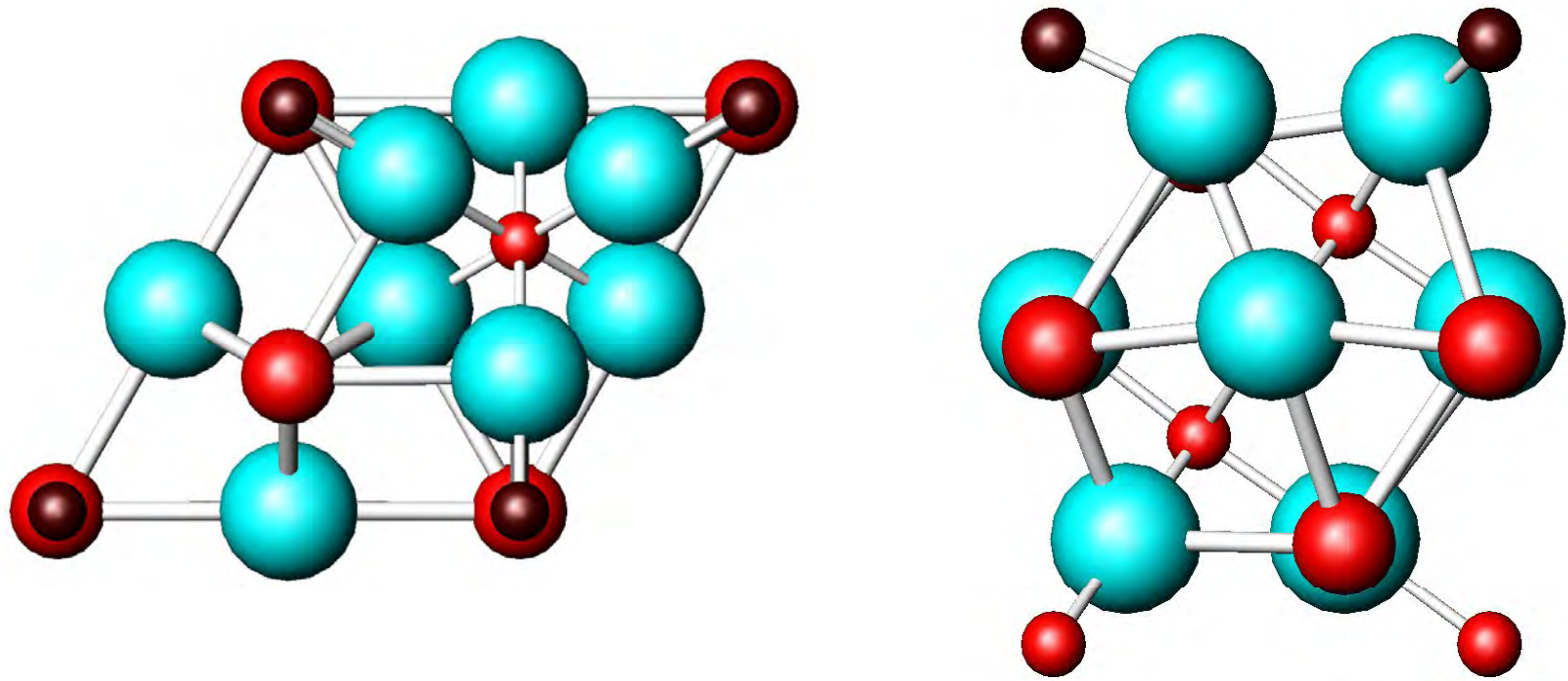

Figure 2a Model 1, non valence-neutral Ti terminated surface in top (left) and side (right). The top Ti atoms are dark brown.

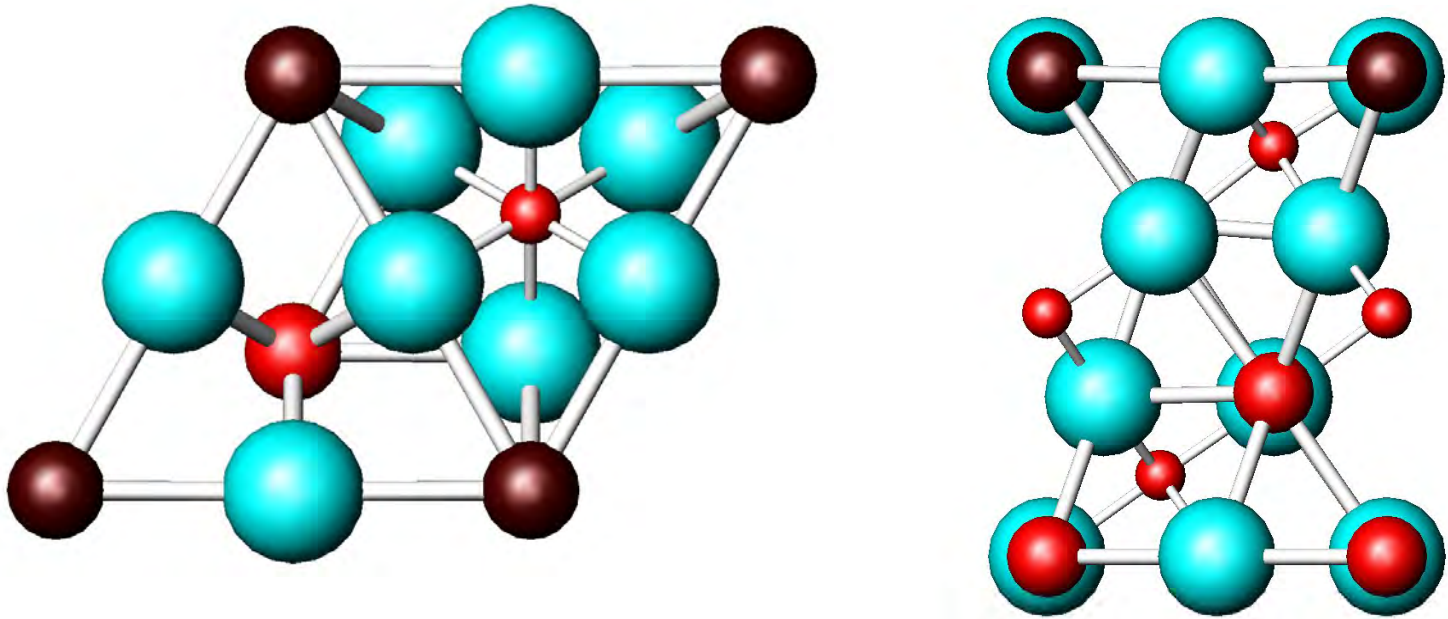

Figure 2b. Model 2, non valence-neutral SrO$_3$ terminated surface, top (left) and side (right). The outermost Sr atoms are dark brown.

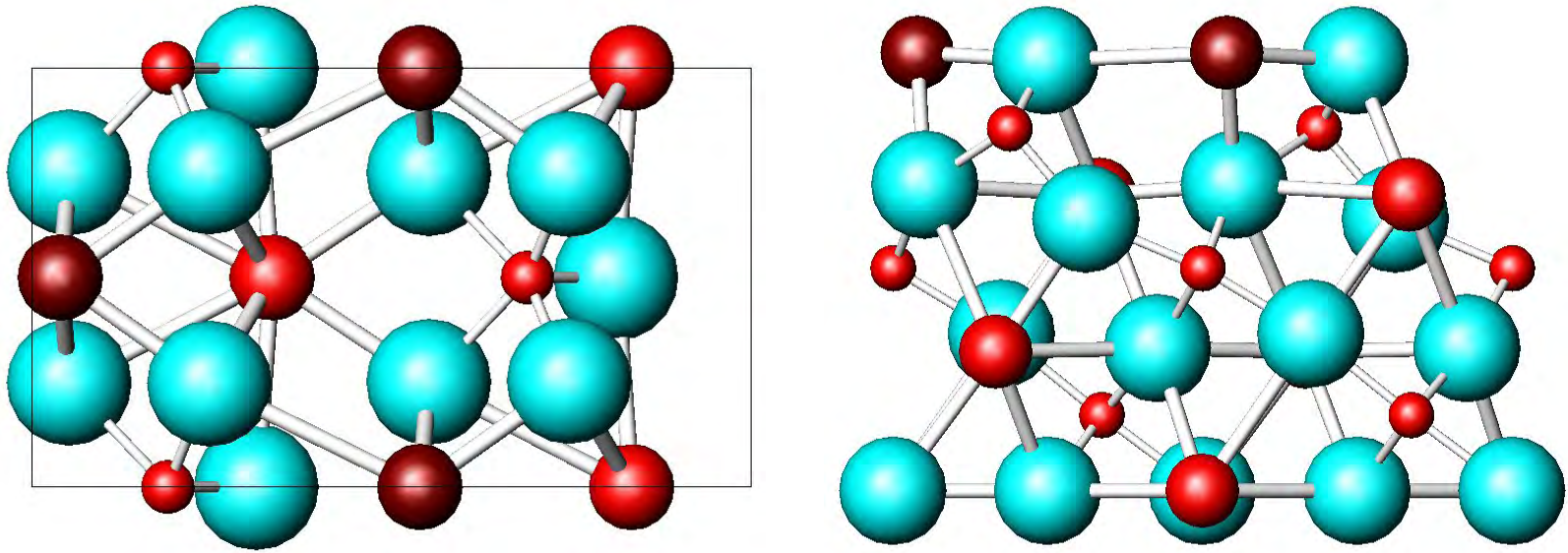

Figure 2c: Top (left) and side (right) view of Model 3, a c2x1 valence compensated surface with 2 O vacancies per surface cell.

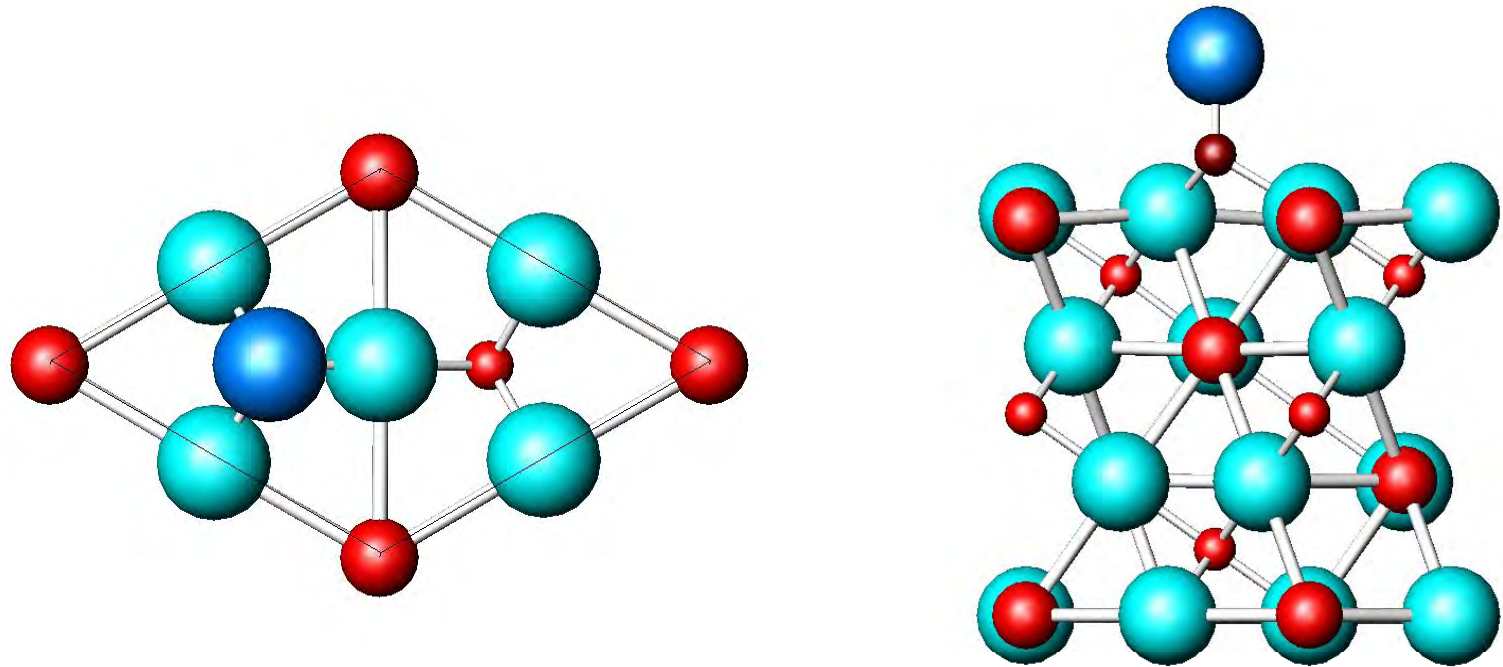

Figure 2d: Top (left) and side (right) view of Model 4, valence compensated 1x1 TiO terminated surface with the additional oxygen atom (dark blue)

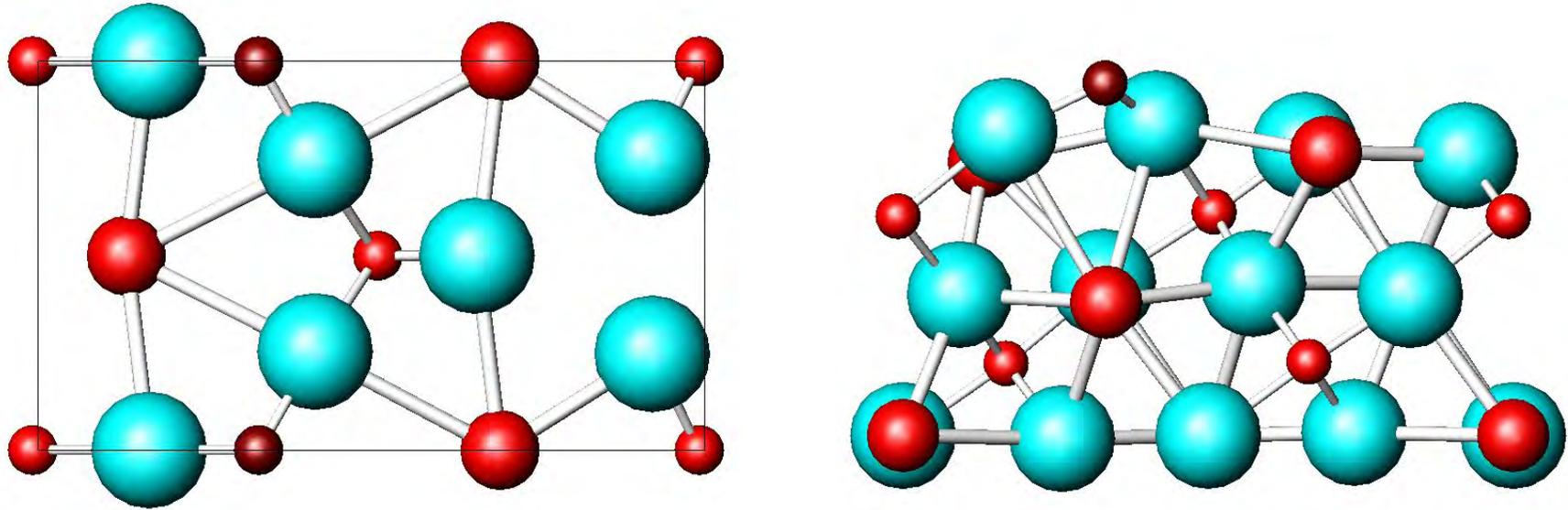

Figure 2e: Top (left) and side (right) view of Model 5, valence compensated 2x1 Ti terminated surface with 2 Ti vacancies per unit cell

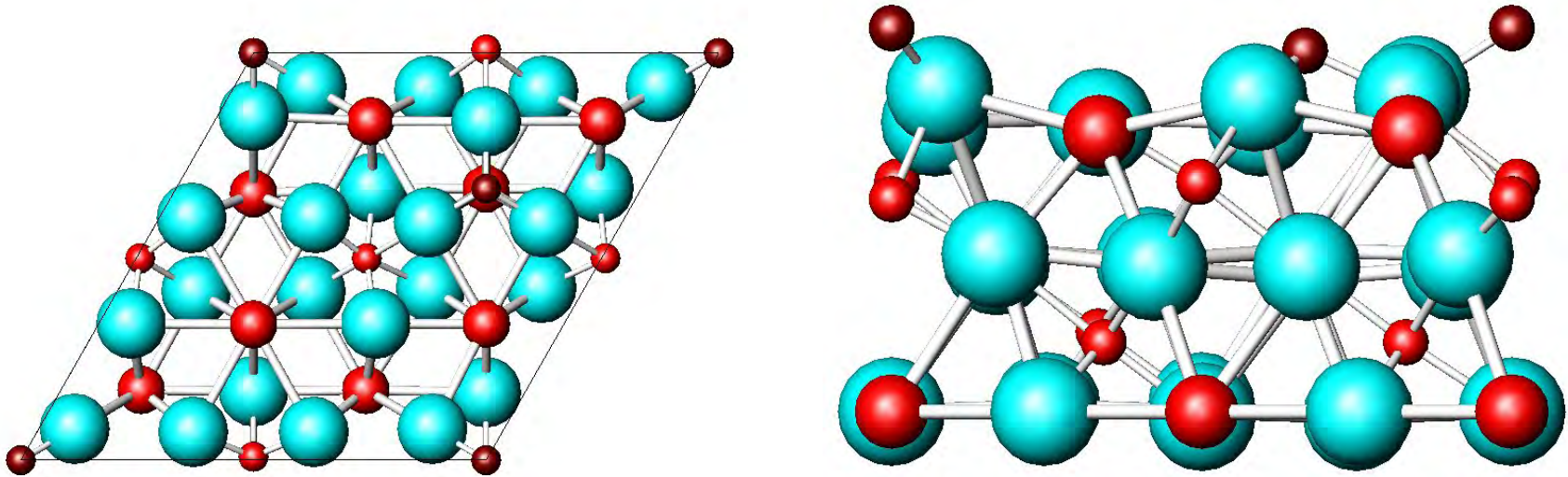

Figure 2f: Top (left) and side (right) view of Model 6, valence compensated Ti terminated surface with 50% Ti vacancies

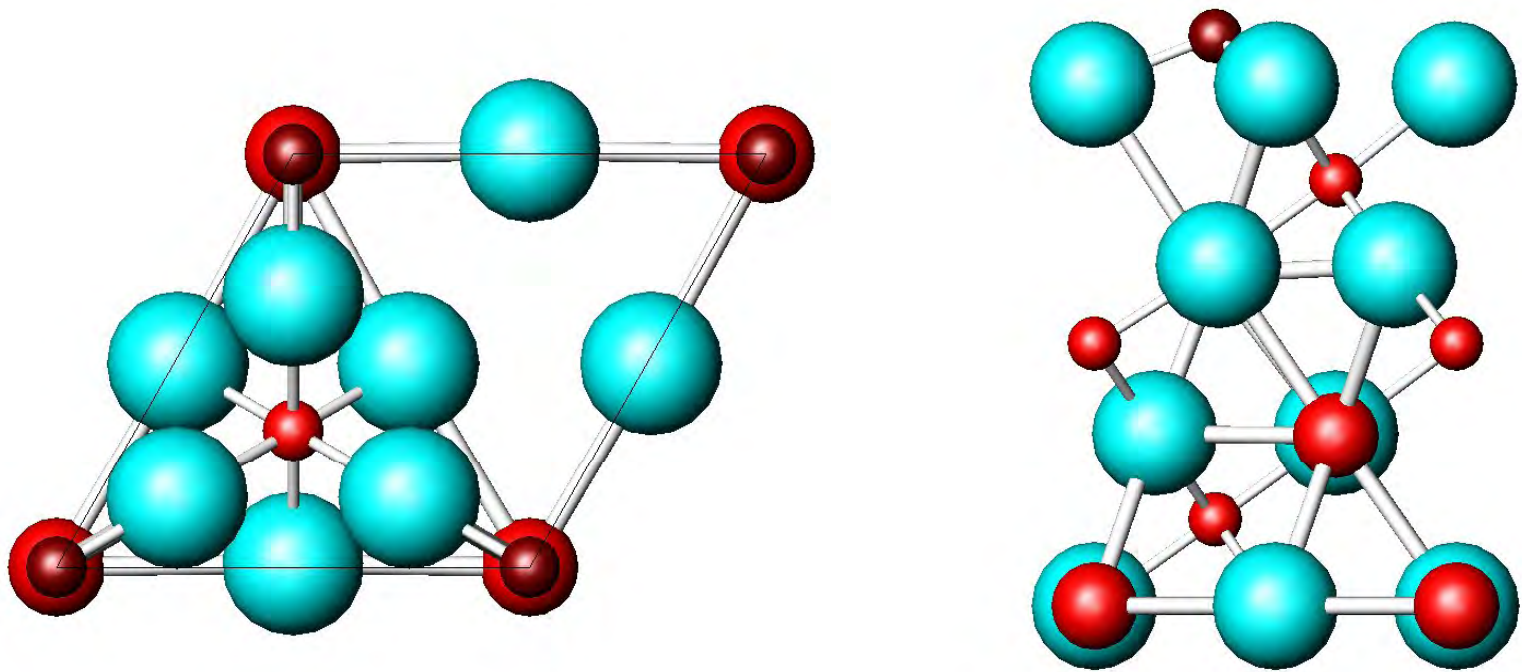

Figure 2g: Top (left) and side (right) view of Model 7, valence neutral 1x1 cell with a Sr vacancy in the second layer

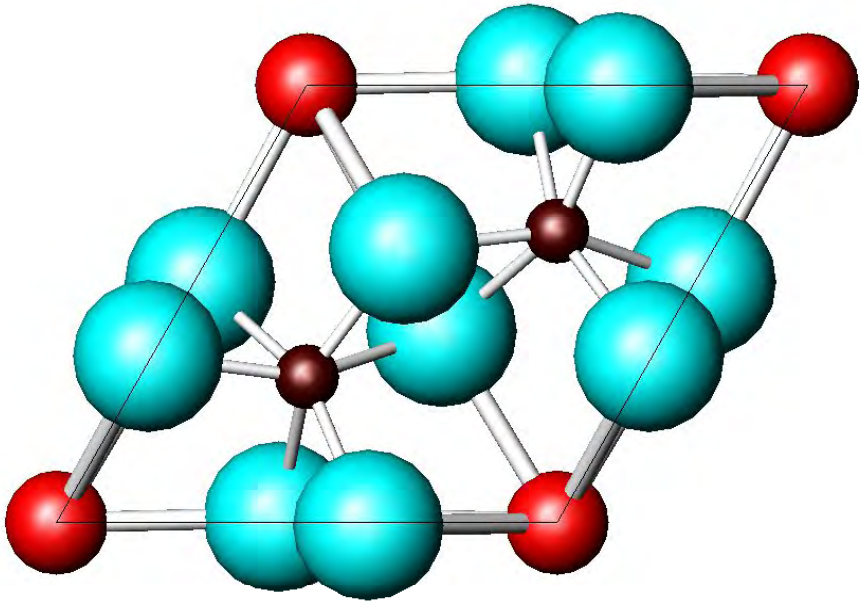 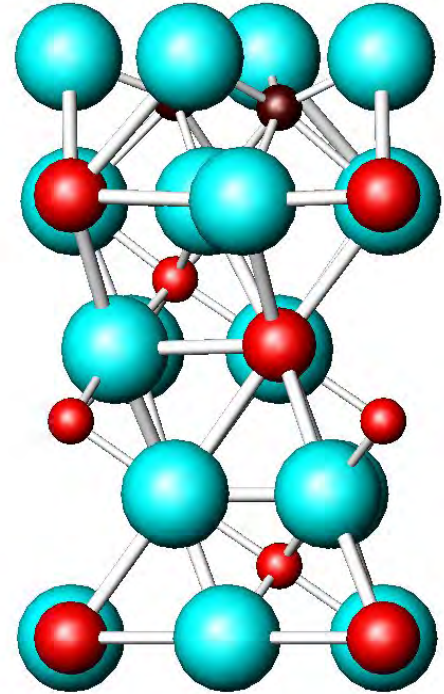

Figure 2h: Top (left) and side (right) view of Model 8, a valence compensated TiO2 rich 1x1 cell with triangular-prismatic co-ordination.

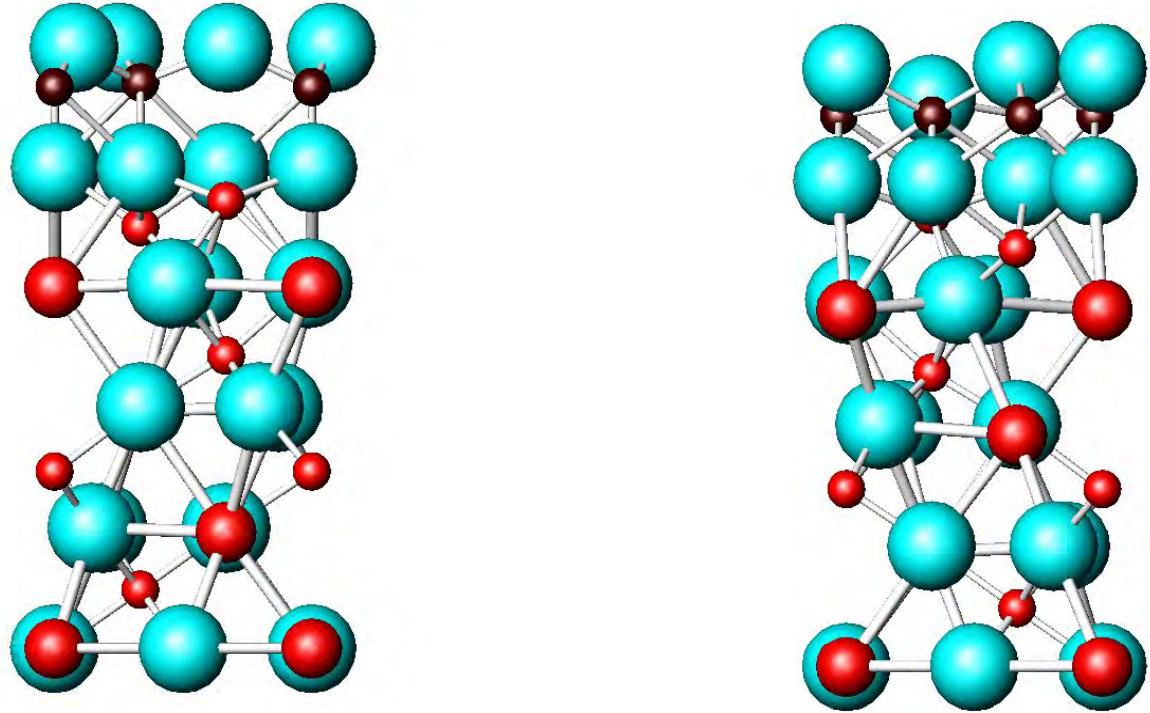

Figure 2i: Side views of Model 9 (left), a Ti2O3 layer on top of Model 8 and a 3(TiO) layer for Model 10 (right)

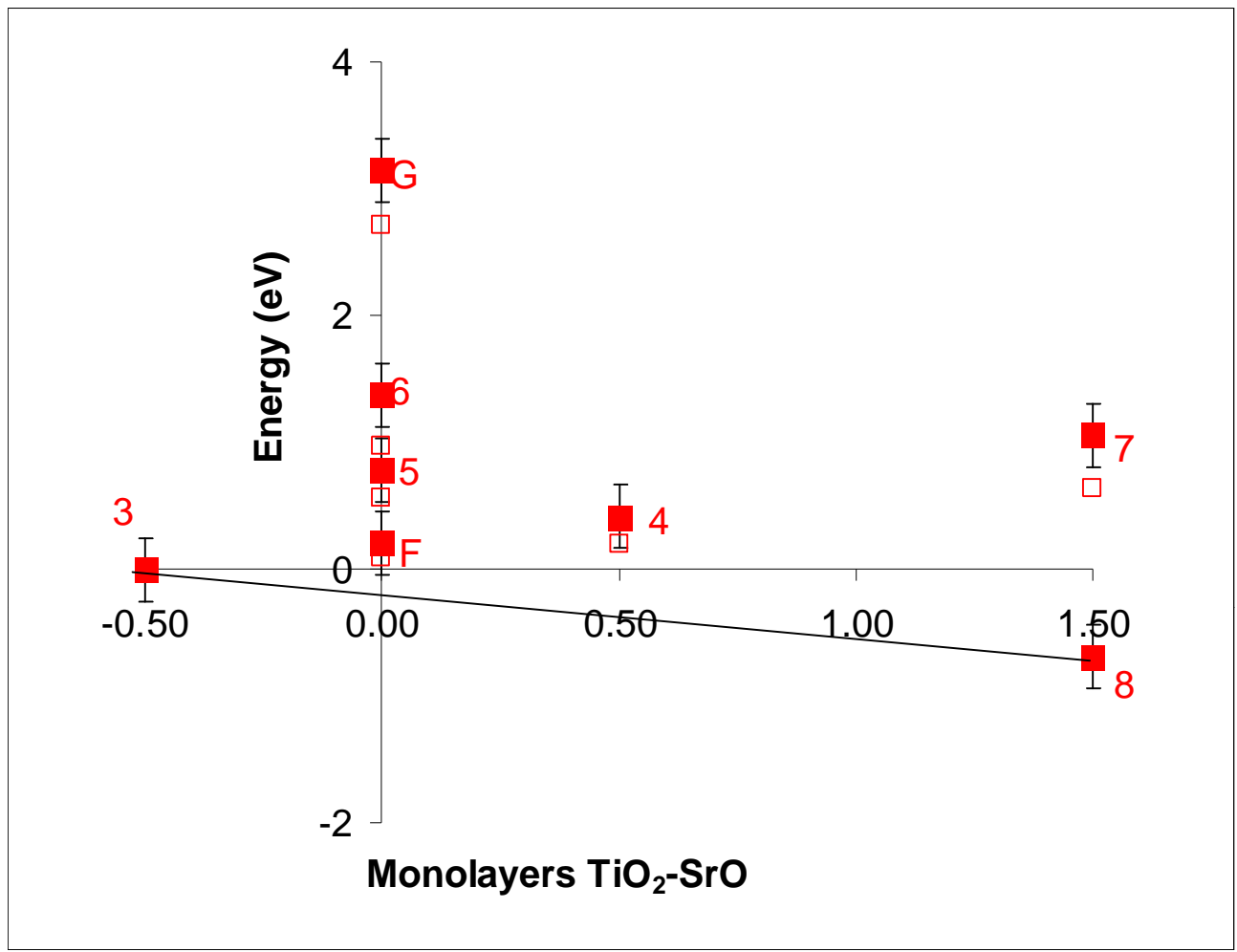

Figure 3: Plot of the energies for the different valence compensated structures relative to the SrO rich Model 3 as labeled, versus stoichiometry with an 0.25eV error bar included, labeled with the relevant model; solid for TPSSh and empty for PBE. Also shown is the fracture surface energy (F), and the Gibbs surface energy (G).

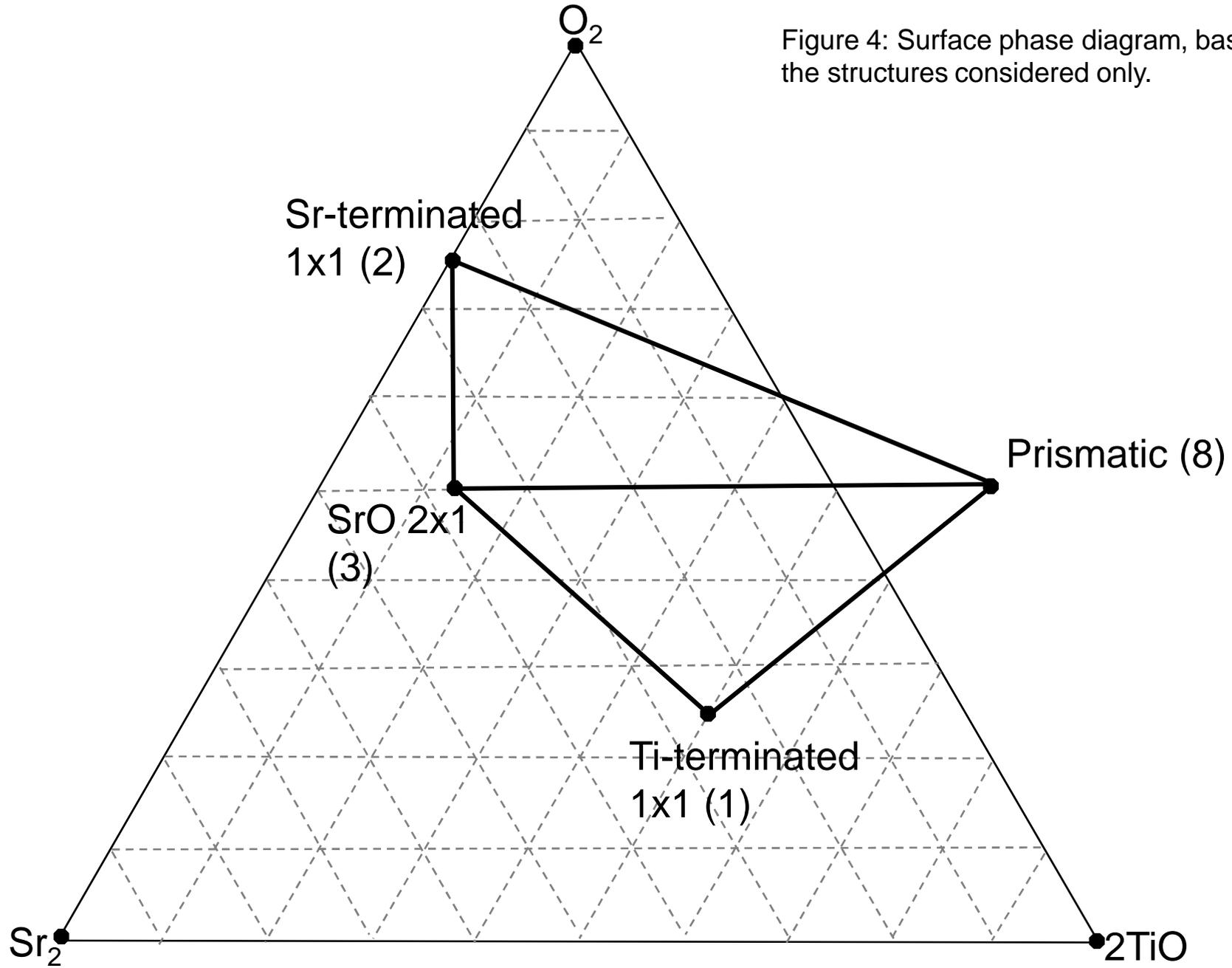

Figure 4: Surface phase diagram, based upon the structures considered only.